\documentclass[a4paper,10pt]{article}
\usepackage{cite}
\usepackage{listings}
\usepackage{graphicx}
\usepackage{url}

\begin{document}
\title{Software Metric Framework}
\author{Charles Hathaway \\ Rensselaer Polytechnic Institute}
\maketitle

				
The Software Metric Framework (SMF) is a design and reference implementation that enables the testing and verification of software metrics on data from publicly available data sources, such as Maven. It describes a pipeline of components and stages that can be adjusted to fit many use cases.

\section{Purpose}

Many researchers have criticized the field of Software Complexity metrics for the lack of testing, verification, and reproducibility of many metrics and case studies that utilized those metrics. This document describes SMF,  a tool that can help address some of these concerns, namely by enabling verification of metrics, reproducibility of experiments, and ease of implementation for new metrics.
The tool in question is the Software Metric Framework; an extensible set of scripts, tools, and standards that allow others to implement metrics in a way that allows automated data collection and analysis. Because it is only a prototype, the framework has been limited to the analysis of Java applications that utilize Maven, a build system which greatly simplifies the task of compiling source code.

\section{Verification}

The first stage in helping this field become more solidified is verifying that existing metrics have some correlation to real-world cases. Although some metrics have been tested, they were usually tested on no more than a couple different applications of varying degrees of complexity. However, most metrics have been proposed with no case study to verify the claims made by authors; sometimes the process by which the authors derive the metrics is more intution than scientific, and the lack of experimental evidence of these metrics make one question their validity.

To help address this problem, the Software Metric Framework (SMF) provides a corpus of data that can be used as a testing bed for hypotheses. This corpus consists of 3 key elements for many selections of applications:

\begin{itemize}
	\item The source code of the application; this is what can be analyzed by a proposed metric
	\item An index of meta data related to the source code, including:
	\item A list of bugs and corresponding changes to the software
	\item A list of versions released and time of release
	\item Meta data related to changes (such as time, author, etc.)
	\item Build artifacts related to the application at a given point in time
    \begin{itemize}
		\item Namely object files, or equivalent
       \end{itemize}
\end{itemize}

This is not distributed with the framework itself, however, the process of recreating the artifacts is automated and documented
The provided corpus is from open source projects hosted by the Apache Foundation. These projects were selected at random from their library, so long as they met the following conditions:

\begin{itemize}
	\item They must have a VCS mirror that SMF can understand; at the time of writing, Git mirrors were the only accepted format
	\item They must make use of Apache’s JIRA installation, which was used to harvest statistics about bugs and versions.
	\item The presence of a pom.xml file in the root directory of the project mirror; this is the toolchain that is used to construct the build artifacts for analysis. 
\end{itemize}

Some types of analyses, especially around meta data, can be done without these build artifacts. To support this pattern, the "--no-compile" flag was added to the run\_metric command. Without this flag, projects and versions that fail to build are skipped

\section{Comparison to related tools}

There are a number of tools out there which do similar things; but they are slightly distinct from SMF.
In the domain of software complexity, where static semantic analysis is a must, Soot \cite{vallee1999soot} and WALA \cite{wala} reign supreme.
SMF does not aim to replace these tools; where they are responsible for calculating a metric, SMF is responsible for aggregating the values they put out and producing meaningful analysis.

A closer synonym is Luigi \cite{luigi}.
This is another tool which aims to automate a pipeline, but where Luigi automates "a pipeline" SMF automates a software metrics pipeline.
One could wrapping SMF in Luigi to provide additional scalability and organization, but for most uses, SMF provides enough resilience that an additional layer of complexity is not needed.

\section{Reproducibility}

\begin{figure}[!htb]
        \caption{Pipeline State Diagram}
        \label{pipeline}
        \centering
        \makebox[\textwidth]{\includegraphics[width=.7\paperwidth]{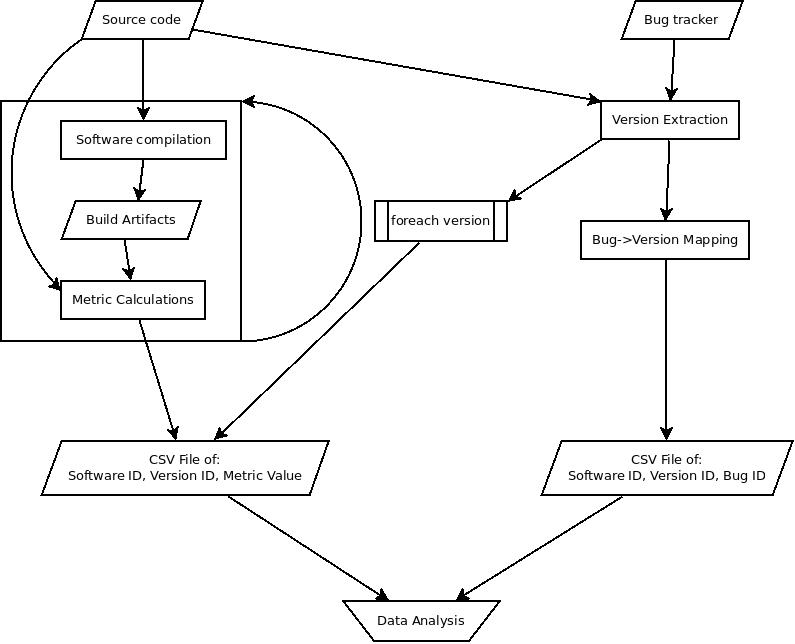}}
\end{figure}

In addition to the ability to verify metrics, it is also essential that the experiments conducted can be verified by a third party. To this end, SMF provides a pipeline of automated scripts, build tools, and data parsing suites which will allow future authors to more easily reproduce an experiment. Figure \ref{pipeline} gives a high level overview of how the pipeline works, and how the various pieces of input come together.

The current state of each of these life cycle stages, and how they are implemented, are described in the next few sections.

\subsection{Pipeline}

The pipeline seen in figure 1 has 2 distinct paths; source code $\rightarrow$ build -> metric calculation $\rightarrow$ CSV $\rightarrow$ analysis (compilation pipeline), and source code + bug tracker $\rightarrow$ version extraction $\rightarrow$ mapping $\rightarrow$ CSV $\rightarrow$ data analysis (metadata pipeline). Each of these pipelines has many intricacies, and will be discussed at length below.

\subsubsection{Compilation Pipeline}

The compilation pipeline is responsible for generating the build artifacts (compiling), which are often used as part of the metric calculation, and doing the calculation, for an individual version. To help simplify this step, we limit the projects in our corpus to projects which utilize Maven; a Java build automation system. The project's source code repository and bug tracker must be manually located prior to this step. It should be entered into the database using the provided web forms under http://127.0.0.1:8000/admin/data\_gathering/project. 

To assist with projects that require extra setup or cleanup steps, three hooks are configurable in the above mentioned web interface. These are:

\begin{itemize}
	\item prebuild\_script: this is called just before the compilation of the project, and should be used if, for example, an additional 		repository needs to be added to the pom.xml file
	\item postbuild\_script: this is called just after the project was compiled, before the metrics are run on the repository.
	\item cleanup\_script: this is called after all the metrics are run, after “mvn clean” is run, but before the next version is checked out
\end{itemize}

Please note that these scripts should not modify the VCS settings (as SMF extensively uses this), nor should they create a situation which will cause metrics to fail.
After being compiled, the metric is calculated. Although most of our metrics are created using Python scripts and Bash commands, there are some exceptions, and so the framework needs to be sufficiently robust to allow for customization while remaining as flexible as possible. To facilitate this, SMF allows the creation of a hook (often an executable bash script) in a folder named like metrics/<language>/<metric suite name>/<executable file>. In addition, it provides a library of common tasks in various languages, described in the folder for that language. The authors of a metric should try and maximize use of the library to avoid writing code that is non-portable.

It is important that each metric not alter any generated files, nor should they take it upon themselves to cleanup after a build. The metrics are run sequentially on the generated artifacts, to save processing time, and assume that they are introduced in a good state.

\subsubsection{Metadata Pipeline}

In addition to compilation, the task of retrieving and indexing data related to a project can be quite daunting. To that end, SMF provides scripts to:

\begin{enumerate}
	\item Download Git repositories for analysis
	\item Convert SVN repositories to Git repositories
	\item Interface with API’s for:
    \begin{enumerate}
		\item JIRA
		\item Bugzilla
    \end{enumerate}
\end{enumerate}

These scripts are written in Python, with extensive usage information that can be found by following the setup instructions (https://github.com/chuck211991/SMF) and running “python manage.py --help”.

The command "fetch\_project" will download a project’s source tree, download a list of bugs along with pieces of meta data, and attempt to match versions in the bug tracker to branches in the source tree.

Since SMF doesn’t store all the metadata by default, due to the sheer quantity of it, we recommend that authors feel free to add fields to the models.py file and modify data\_gathering/management/commands/fetch\_project.py as needed. Using Django allows us to utilize tools that have extensive documentation and plenty of examples online. Once you start putting your changes in, please send a pull request so we can incorporate them into the main repository. 

\section{Usage}

To obtain SMF, please visit the software home page at \url{https://gitlab.com/charles.hathaway/software-metrics-framework}.

To assist new users in using this tool, it is required that all scripts include a help section. Below is a summary of the help available from the most important commands. The most important files are:

\subsection{manage.py}

This includes tasks to fetch projects and run metrics; aptly named fetch\_project and run\_metric.
It also is the tool we use to load and dump data in a way that can be later recalled regardless of the backend database we are using
metric/interactive\_complexity/*

These are various scripts that demonstrate how we might create a metric; you can use any language that the shell will understand and know how to execute. This gives us a great deal of flexibility for future 

\begin{lstlisting}
% python manage.py run_metric -h
usage: manage.py run_metric [-h] [--version] [-v {0,1,2,3}]
                            [--settings SETTINGS] [--pythonpath PYTHONPATH]
                            [--traceback] [--no-color]
                            [--git-repo-base REPO_BASE] [--no-compile]
                            [--project PROJECT]
                            shell_script [shell_script ...]

Runs the script specified as an argument, passing in the path to MVN project
as an argument The script should can report metric values by printing to
STDOUT like this: #>> METRIC_NAME=METRIC_VALUE The output is grepped for #>>,
then the values split and stored If the script runs longer than --timeout
(default, 5 minutes) it gets killed

positional arguments:
  shell_script          A script to run

optional arguments:
  -h, --help            show this help message and exit
  --version             show program's version number and exit
  -v {0,1,2,3}, --verbosity {0,1,2,3}
                        Verbosity level; 0=minimal output, 1=normal output,
                        2=verbose output, 3=very verbose output
  --settings SETTINGS   The Python path to a settings module, e.g.
                        "myproject.settings.main". If this isn't provided, the
                        DJANGO_SETTINGS_MODULE environment variable will be
                        used.
  --pythonpath PYTHONPATH
                        A directory to add to the Python path, e.g.
                        "/home/djangoprojects/myproject".
  --traceback           Raise on CommandError exceptions
  --no-color            Don't colorize the command output.
  --git-repo-base REPO_BASE
                        Where to store the Git repos that must be downloaded
  --no-compile          Do not compile projects before calling the metrics
  --project PROJECT     Which project to run on; if omitted, runs on all
                        projects
\end{lstlisting}

\section{Adding a New Metric}

Most metrics will require a fair amount of novel programming; we do not document that step here. Instead, we will demonstrate writing a Python script which utilizes some of the provided libraries to speed development. This Python script can be seen in the SMF library itself under metrics/python/interactive\_complexity/rfc.py.

NOTE: we don’t include setting up SMF in these instructions; that is the responsibility of the reader. The environment in which this tools runs is heavily influenced by the scale and performance requirements of the user, and so we don’t say much. There is a Dockerfile and docker-compose.yml which can help one get started with SMF; if you use this method, please alter the Bash commands below to use “docker-compose run” and appropriate paths.

\subsection{Skeleton}

Most Python scripts will begin with something like this:

\begin{lstlisting}
#!/usr/bin/env python

file_description = """
Calculated the IC-RFC metric for a given project
"""
import argparse
import doctest
import os
import logging

def get_arg_parser():
    parser = argparse.ArgumentParser(description=file_description)
    parser.add_argument('--test', action='store_true', dest='test',
        default=False, help='Runs the built in tests')
    parser.add_argument("-v", "--verbose", help="increase output verbosity",
        action="store_true")
    parser.add_argument('project_root', type=str,
        help='location of the project directory')
    return parser

def main(project_root, **arguments):
    logging.info("Scanning POM...")
    pass

if __name__ == "__main__":
    parser = get_arg_parser()
    parser = vars(parser.parse_args())
    if parser['verbose']:
        logging.basicConfig(level=logging.DEBUG)
    if parser["test"] == True:
        doctest.testmod()
    if os.path.isdir(parser["project_root"]):
        parser["project_root"] = os.path.join(parser["project_root"], "pom.xml")
    main(**parser)
\end{lstlisting}

The important elements to consider when making your script:
\begin{enumerate}
	\item Documentation. This file has a description, and adheres to the standard that you support the --help flag
	\item Usage of "logging" instead of just printing things. This helps to manage output when we run many metrics at once
	\item The presence of "\#!/usr/bin/env python" at the top; this script is usually executed as a generic executable, which means that 		the interpreter should be specified at the top of the file
\end{enumerate}

\subsection{Using Java Tools}

Now that we have a basic script, let’s make it do something. Interactive Complexity defines Response for Class as:
the number of unique methods calls exposed to applications that depend on this one.
And one easy way to get that number is by calling the “javap” command on each class file, then counting the number of function calls declared as public. This can be broken down to two simple steps:
\begin{enumerate}
	\item Get the list of precompiled class files
			This is done by the pipeline, and to avoid redoing work, we should not recompile it ourselves
	\item For each file in that list, scan it using javap and search for the public keyword
\end{enumerate}
We alter the main function to look this:

\begin{lstlisting}
import pipes
from data_gathering.library.commands import execute_command
...
def main(project_root, **arguments):
    logging.info("Scanning POM...")
    total = 0
    dir = os.path.dirname(project_root)
    class_files = [line for line in execute_command("find %s -name *.class" % dir)]
    for c in class_files:
        c = c.strip()
        result = execute_command("javap %s" % (pipes.quote(c)))
        count = 0
        for r in result:
            r = r.strip()
            if r.startswith("public"):
                count += 1
        if count > 0:
            total += count - 1 # minus one for class declaration
    print "#>> IC-RFC=%s" % total
    return total
\end{lstlisting}

Note the use of the library helper, execute\_command. This function takes a command as an argument, and yields the results line by line as queried. It will block until a line becomes available, or the application terminates. This is a wrapper for the Python subprocess module.

It’s very important that metric output "\#>> SOMEMETRIC=42", with SOMEMETRIC replaced by the name of your metric and the number (42) being replaced by the value of your calculation.

\subsection{Testing the Metric}

Once your metric is created, you need to test it before you call it on all metrics. We try to limit the amount of entropy in the corpus, but depending on your metric, multiple scenarios may need to be tested before you invest a great deal of time in running it on all projects. We simply test the above on a single versions of a single project.

\begin{lstlisting}
(python) chathaway@blaze:~/programming/data_gathering$ chmod +x test.py 
(python) chathaway@blaze:~/programming/data_gathering$ python manage.py run_metric --project HTTPCLIENT --sha "3aeb27" ./test.py 
\end{lstlisting}
The script will run for a while, and hopefully eventually output "\#>> IC-RFC=8690.00000", or something along those lines.

\bibliographystyle{plain}
\bibliography{main}

\end{document}